\documentclass[floatfix,aps,amsmath,nofootinbib,onecolumn,10pt]{revtex4}
\pdfoutput=1

\usepackage{graphicx}
\usepackage{bm}

\def\({\left(}
\def\){\right)}
\def\[{\left[}
\def\]{\right]}

\def\e{\begin{equation}}
\def\q{\end{equation}}
\def\m{\begin{eqnarray}}
\def\n{\end{eqnarray}}

\begin{document}


\title{Effect of the Early Reionization on the Cosmic Microwave Background and Cosmological Parameter Estimates}

\author{Qing-Guo Huang\footnote{huangqg@itp.ac.cn} and Ke Wang \footnote{wangke@itp.ac.cn}}
\affiliation{$^1$CAS Key Laboratory of Theoretical Physics, Institute of Theoretical Physics, Chinese Academy of Sciences, Beijing 100190, China\\
$^2$School of Physical Sciences, University of Chinese Academy of Sciences, No. 19A Yuquan Road, Beijing 100049, China}

\date{\today}

\begin{abstract}

The early reionization (ERE) is supposed to be a physical process which happens after recombination, but before the instantaneous reionization caused by the first generation of stars. We investigate the effect of the ERE on the temperature and polarization power spectra of cosmic microwave background (CMB), and adopt principal components analysis (PCA) to model-independently reconstruct the ionization history during the ERE. In addition, we also discuss how the ERE affects the cosmological parameter estimates, and find that the ERE does not impose any significant influences on the tensor-to-scalar ratio $r$ and the neutrino mass at the sensitivities of current experiments. The better CMB polarization data can be used to give a tighter constraint on the ERE and might be important for more precisely constraining cosmological parameters in the future.

\end{abstract}

\pacs{????}

\maketitle


\section{Introduction}

According to the standard model of cosmology, the universe was almost full of neutral hydrogen and helium after the epoch of recombination. However, many direct measurements of the ionization state of the universe including the Gunn-Peterson effect in QSOs \cite{Fan:2005es,McGreer:2014qwa,Schroeder:2012uy} and Lyman alpha emission in galaxies \cite{Pentericci:2014nia,Schenker:2014tda,Tilvi:2014oia} indicate that intergalactic gas has been almost fully reionized by $z\sim6$ and the universe was no longer neutral at $z<10$. But we don't know when the transition, so-called cosmic reionization, took place. In the literature there are many different candidates for the sources of the cosmic reionization for different onset of the reionization process. Although star-forming galaxies at $6\lesssim z<10$ are taken as the main agents of reionization \cite{Robertson:2013bq,Robertson:2015uda}, Thomson optical depth to electron scattering $\tau_\textrm{re}$ derived from star-forming rate $\rho_\textrm{SFR}$ is smaller than that constrained by Planck \cite{Ade:2015xua}.
On the other hand, much higher redshift ionization sources are still allowed. See, for example, \cite{Heinrich:2016ojb,Miranda:2016trf}. 
There are several possible processes that might had modified the ionization state of the universe at high redshifts: the decay or annihilation of dark matter (DM) transmuted mass of DM into energy, a fraction of which would be deposited into the intergalactic medium (IGM) and then heat, ionize or excite the neutral atoms \cite{Slatyer:2009yq,Chluba:2009uv,Finkbeiner:2011dx,Liu:2016cnk,Oldengott:2016yjc}; primordial black holes (PBHs) immersed in an baryon gas will be accreted onto by gas and DM, which produced radiation heating, ionizing or exciting the IGM \cite{Ricotti:2007au,Chen:2016pud,Ali-Haimoud:2016mbv}.
In fact, not only can a well-understood ionization history of the universe help us to probe the microscopic properties of DM and the abundance of massive PBHs in turn, but also is important for determining cosmological neutrino mass \cite{Smith:2006nk,Allison:2015qca}, detecting CMB B-Modes from inflationary gravitational waves \cite{ Kamionkowski:2015yta}, exploring the large scale anomalies in the CMB \cite{ Mortonson:2009xk,Mortonson:2009qv} and testing the single-field slow-roll consistency relation \cite{ Mortonson:2007tb}.

In this paper we investigate the early reionization (ERE) epoch between the standard instantaneous reionization caused by the first generation of stars and recombination. Here we don't specify what physical mechanism causes it, but turn to a model-independent method, called principal component analysis (PCA), to reconstruct the ionization history during the ERE. Although applying PCA to reionization have been done in many works, for instance \cite{Hu:2003gh,Dai:2015dwa,Liu:2015gho}, they only explored the low-redshift region ($z<30$). Here we focus on the ERE which may happen in the region of $20\lesssim z<910$, and see how the ERE affects the cosmological parameter estimates.

This paper is organized as follows. In Sec.~\ref{cmb}, we sketch out the effects of ERE on the CMB temperature and polarization power spectra. In Sec.~\ref{pca}, we utilize PCA to model-independently reconstruct the ionization history during the ERE from Planck 2015 data. In Sec.~\ref{fitting}, we investigate how the ERE affects the estimates of the tensor-to-scalar ratio and the neutrino mass. Summery and discussion are given in Sec.~\ref{sd}.

\section{Effects of the Early Reionization on the CMB power spectra}
\label{cmb}

The recombination of Helium III to Helium II, Helium II to Helium I and Hydrogen recombination last from $z=10^4$ to late time, and the ionization fraction defined by \begin{equation}
x_{\textrm{e}}(z)\equiv {n_\text{e}\over n_\text{H}}
\end{equation}
decreased from $1.16$ to around $10^{-4}$, where $n_\text{e}$ is the number density of free electrons and $n_\text{H}$ is the total number density of Hydrogen nuclei. When the first generation of early star-forming galaxies were formed, the reionization of neutral Hydrogen and the first reionization of neutral Helium occur in the intergalactic medium and the ionization fraction is usually supposed to be with a $\tanh$-like increase, namely the instantaneous reionization,
\begin{align}\label{}
{x}_\textrm{e}(z)=
\begin{cases}
x_{\textrm{e,rec}}(z), ~~\textrm{for}\ z\geq z_{\textrm{beg}}\ ;\\
\frac{f-{x}_\textrm{e}(z_{\textrm{beg}})}{2}\[1+\tanh\(\frac{y(z_{\textrm{re}})-y(z)}{\Delta _y}\)\]+ {x}_\textrm{e}(z_{\textrm{beg}}), ~~\textrm{for}\ z<z_{\textrm{beg}}\ ,
\end{cases}
\end{align}
where $f=1+{n_{\textrm{He}}}/{n_\textrm{H}}=1.08$ denotes the ionization fraction of a fully ionized universe, $z_\textrm{re}$ is the redshift when the universe is half reionized, $z_\textrm{beg}=z_\textrm{re}+8\times\Delta_z$, $\Delta_y=1.5\sqrt{1+z_\textrm{re}}\Delta _z$ with $\Delta_z=0.5$, $y(z)=(1+z)^{\frac{3}{2}}$ is used by \textsf{CAMB} \cite{Lewis:2008wr}, and $x_\textrm{e,rec}(z)$ is the ionization fraction of recombination history given by \textsf{RECFAST} \cite{Seager:1999bc}. The ERE is supposed to happen before the instantaneous reionzation, and then the ionization fraction for $z\geq z_\text{beg}$ should be modified to $x_{\textrm{e,rec}}(z)+\Delta x_\textrm{e}(z)$, where $\Delta x_\textrm{e}(z)$ encodes the information about the ERE. Therefore, after considering the ERE, the ionization fraction takes the form
\begin{align}\label{eq:xes}
x_\textrm{e}(z)=
\begin{cases}
x_{\textrm{e,rec}}(z)+\Delta x_\textrm{e}(z), ~~\textrm{for}\ z\geq z_{\textrm{beg}}\ ;\\
\frac{f-x_\textrm{e}(z_{\textrm{beg}})}{2}\[1+\tanh\(\frac{y(z_{\textrm{re}})-y(z)}{\Delta _y}\)\]+ x_\textrm{e}(z_{\textrm{beg}}), ~~\textrm{for}\ z<z_{\textrm{beg}}\ .
\end{cases}
\end{align}
The optical depth for Thomson scattering due to the late-time instantaneous reionization is defined by
\begin{equation}
\label{}
\tau_{\textrm{re}}\equiv\int^{z_{\textrm{beg}}}_0\left[x_\textrm{e}(z)-x_{\textrm{e,rec}}(z)\right]n_\textrm{H}(z) \sigma_T \frac{dz}{H(1+z)},
\end{equation}
and the optical depth contributed by the ERE is
\begin{equation}
\label{}
\Delta\tau\equiv\int_{z_{\textrm{beg}}}^{z_*}\Delta x_\textrm{e}(z)n_\textrm{H}(z) \sigma_T \frac{dz}{H(1+z)},
\end{equation}
where $\sigma_T$ is the Thomson cross section and $z_*$ is the redshift of recombination.

The effects on the CMB angular power spectra from the ERE and reionization are encoded in the the photon transfer functions $\Delta_{Tl}^{(S)}(k)$ and $\Delta_{El}^{(S)}(k)$ which are obtained by integrating their corresponding source function $S_{T,E}^{(S)}(k,\eta)$ and spherical Bessel function $j_l[k(\eta_0-\eta)]$ along the line of sight, \cite{Seljak:1996is,Zaldarriaga:1996xe},
\begin{eqnarray}
\Delta_{Tl}^{(S)}(k)&=&\int_0^{\eta_0}d\eta S_T^{(S)}(k,\eta)j_l[k(\eta_0-\eta)], \\
S_T^{(S)}(k,\eta)&=&g\(\Delta_{T0}^{(S)}+2\dot{\alpha}+\frac{\dot{v_b}}{k}+\frac{\Pi}{4}+\frac{3\ddot{\Pi}}{4k^2}\)\\ \nonumber
&+&e^{-\tau}(\dot{\kappa}+\ddot{\alpha})+\dot{g}\(\alpha+\frac{v_b}{k}+\frac{3\dot{\Pi}}{4k^2}\)+\frac{3\ddot{g}\Pi}{4k^2},\\
\Delta_{El}^{(S)}(k)&=&\sqrt{\frac{(l+2)!}{(l-2)!}}\int_0^{\eta_0}d\eta S_E^{(S)}(k,\eta)j_l[k(\eta_0-\eta)], \\
S_E^{(S)}(k,\eta)&=&\frac{3g\Pi}{4(\eta_0-\eta)^2k^2},\\
\Pi&=&\Delta_{T2}^{(S)}+\Delta_{P2}^{(S)}+\Delta_{P0}^{(S)},\\
g&=&-\dot{\tau}e^{-\tau},\\
\tau(\eta)&=&\int_\eta^{\eta_0}d\eta n_\textrm{e}\sigma_Ta\\
\alpha&=&\frac{\dot{h}+6\dot{\kappa}}{2k^2},
\end{eqnarray}
where $h$ and $\kappa$ are the scalar perturbations of metric in the synchronous gauge, $v_b$ is the baryon velocity, $\eta_0$ is the conformal time at present, and the dots denote the derivatives with respect to the conformal time $\eta$. Since the integral $\int_0^{\eta_0}d\eta g(\eta)=1$, the visibility function $g$ is taken as a probability density that a photon last scattered at $\eta$.

For the temperature power spectrum, the source function $S_T^{(S)}(k,\eta)$ consists of three parts: the anisotropy terms with a factor of $g$, the integrated Sachs-Wolfe (ISW) term $e^{-\tau}(\dot{\kappa}+\ddot{\alpha})$ and the anisotropy terms with derivatives of $g$. Since the $x_\text{e}$ evolves smoothly, the anisotropy terms with derivatives of $g$ should not be dominant. The contribution of the ISW effect to the temperature power spectrum is not affected by the modification of the ionization history because the early ISW effect takes place quite before the recombination and the late ISW effect occurs at low redshifts where our universe has been fully reionized. Therefore, the effect of the reionization, including the late instantaneous reionization and the ERE, on the temperature power spectrum mainly comes from the anisotropy terms with a factor of $g$. At recombination, only monopole makes contribution to the temperature power spectrum on large scales, but all of monopole, dipole and quadrupole contribute to the the temperature power spectrum on small scales. After recombination, even though higher multipoles enter horizon on the intermediate and large scales gradually, their contributions are negligibly small. However, a photon might been scattered by the free electrons due to the reionization. Thus, the temperature power spectrum observed today is suppressed by $\int_0^{\eta_*} d\eta g(\eta)\sim e^{-(\tau_{\textrm{re}}+\Delta\tau)}$.

Since the polarization power spectrum on small scales was also formed at recombination, it is suppressed by $\sim e^{-(\tau_{\textrm{re}}+\Delta\tau)}/(\eta_0-\eta_*)^2$. However, the quadrupoles scattered by free electrons through Thomson scattering can induce polarization on the intermediate and large scales when they enter horizon gradually after recombination. If the ERE occurs at around $\eta_\text{ere}$, the polarization power spectrum on the intermediate scales is enhanced by $\sim (e^{-\tau_{\textrm{re}}}-e^{-(\tau_{\textrm{re}}+\Delta\tau)})/(\eta_0-\eta_\text{ere})^2\sim \Delta\tau/(\eta_0-\eta_\text{ere})^2$. Similarly, on the largest scales ($\ell \lesssim 10$), the late instantaneous reionization enhances the  polarization power spectrum by $\sim (1-e^{-\tau_{\textrm{re}}})/(\eta_0-\eta_\text{re})^2\sim \tau_\text{re}/(\eta_0-\eta_\text{re})^2$.

\begin{figure}[!htb]
\begin{center}
\includegraphics[width=9cm]{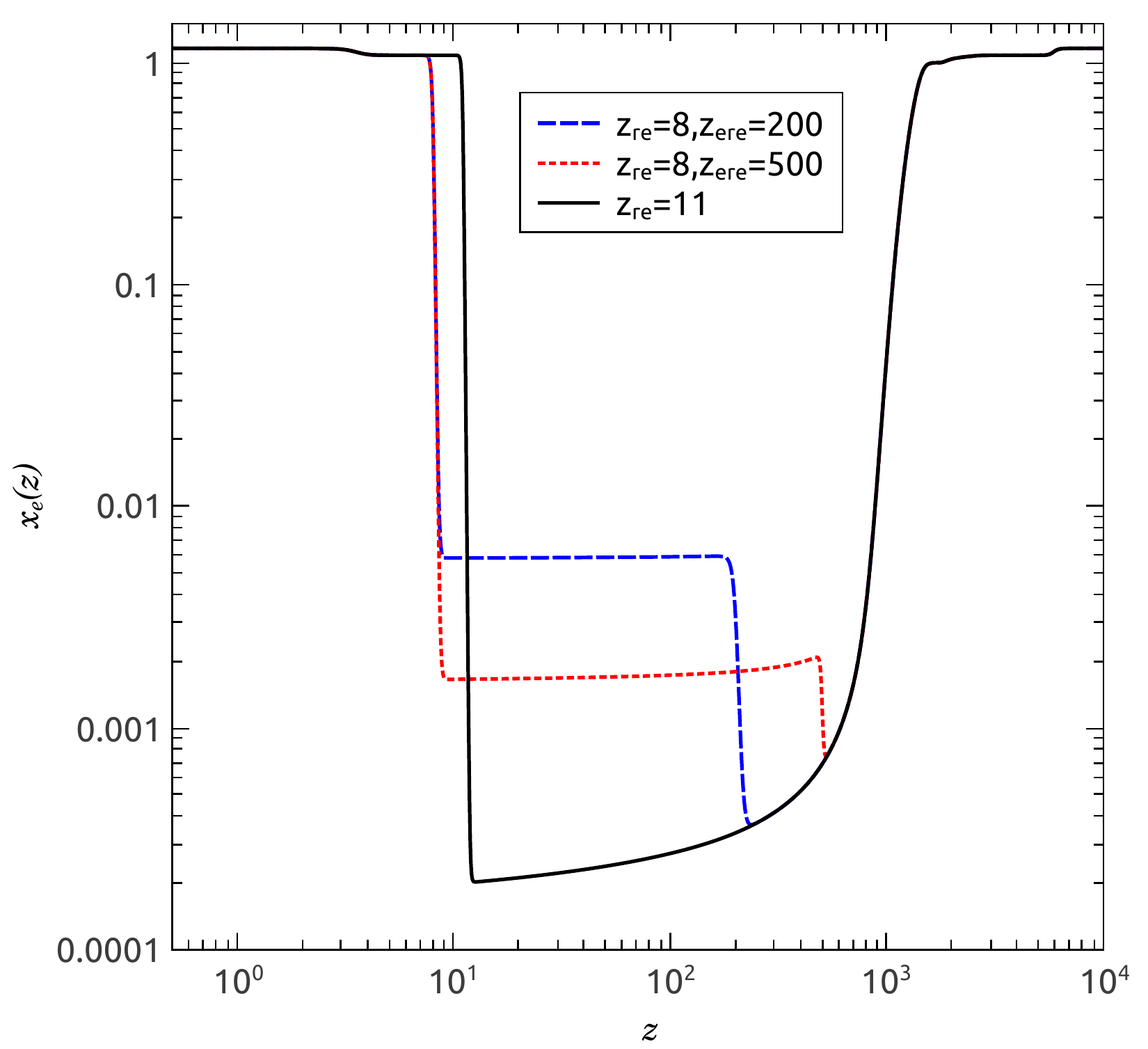}
\end{center}
\caption{Ionization histories of the universe. Here $\tau_{\textrm{re}}+\Delta \tau=0.089$ are kept fixed for these three cases. }
\label{fig:xetanh}
\end{figure}

\begin{figure}[!htb]
\begin{center}
\includegraphics[width=8cm]{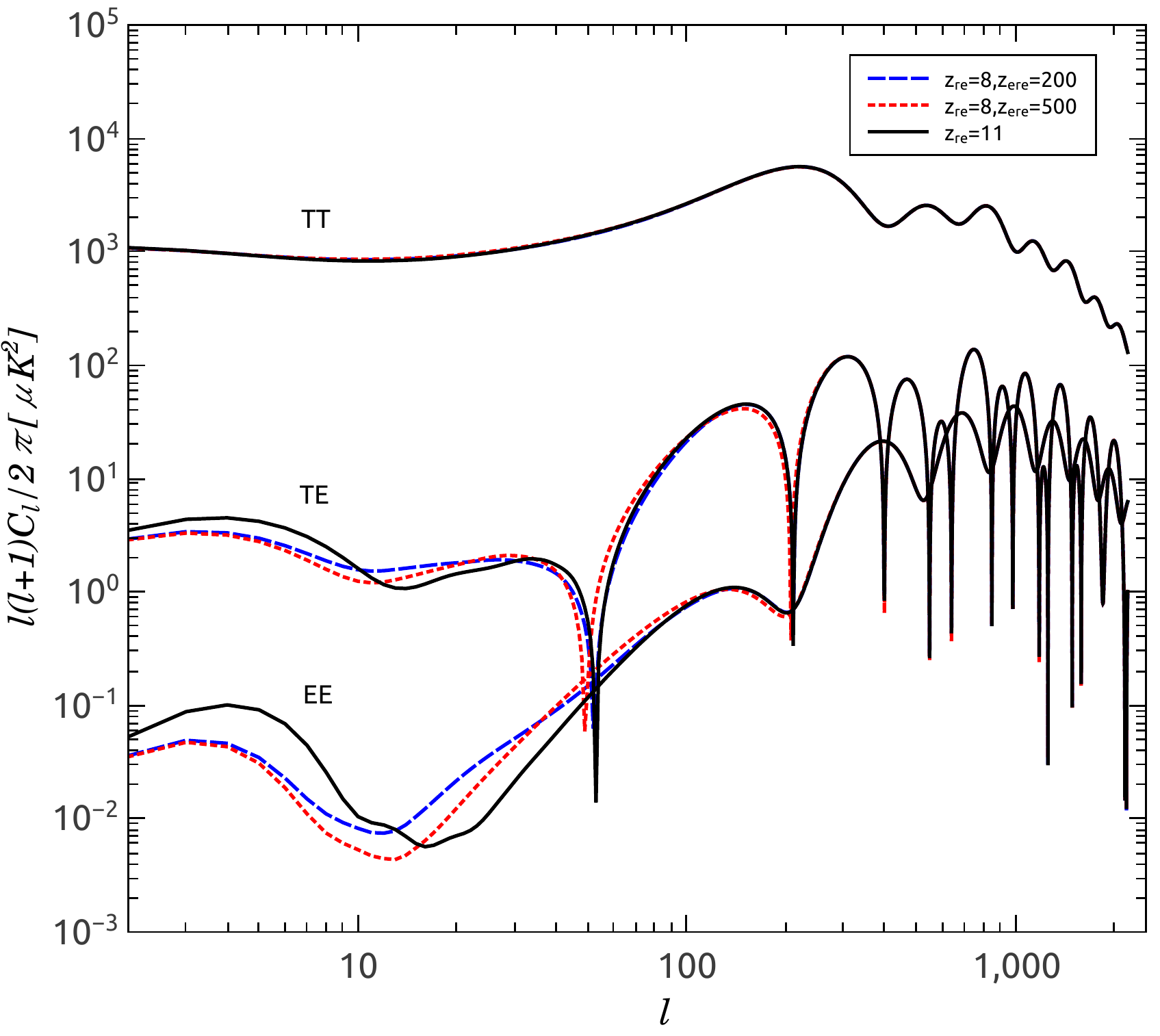}\quad \quad  \includegraphics[width=8.55cm]{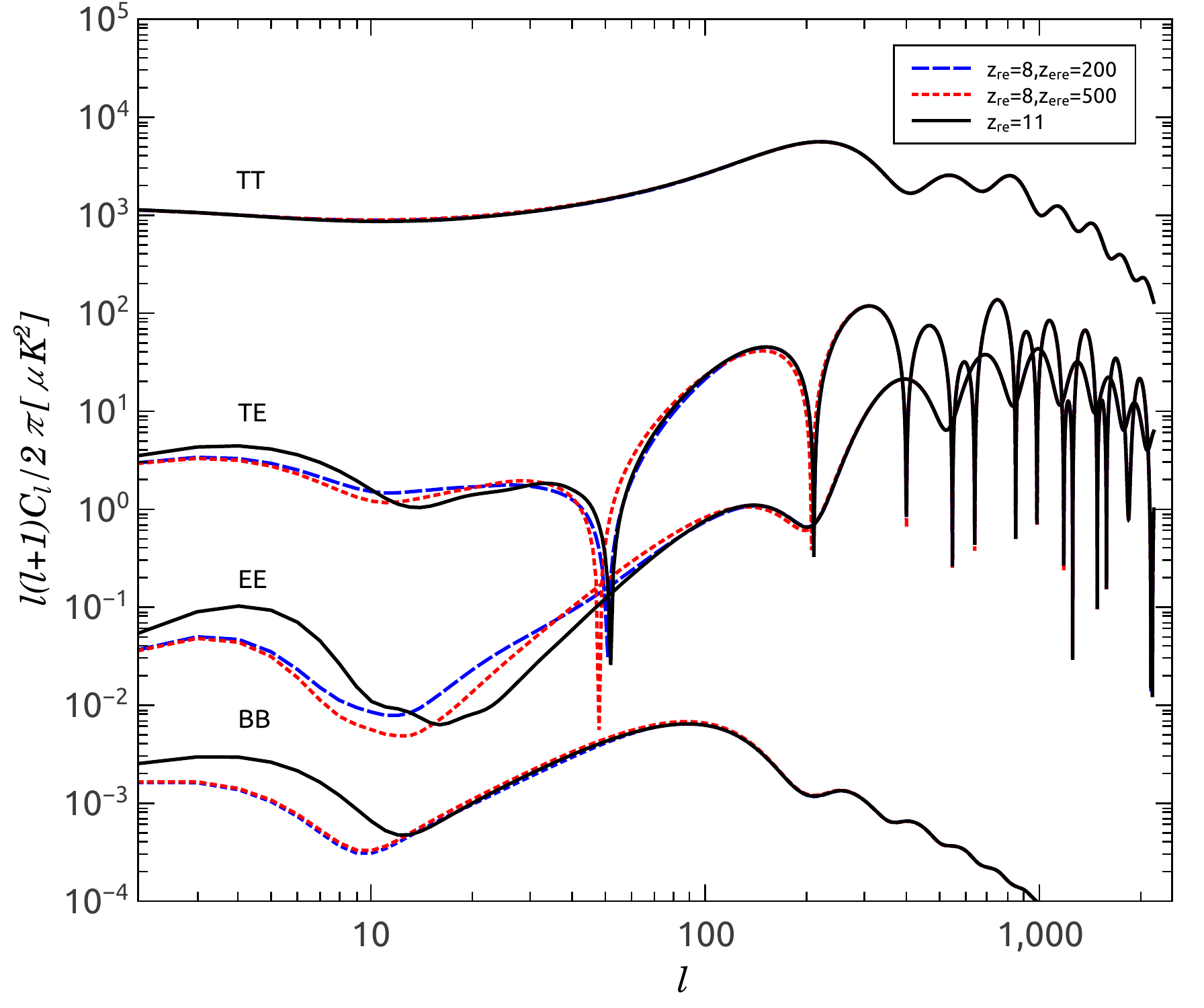}
\end{center}
\caption{Effects on the CMB angular power spectra for different reionization histories. Here the tensor-to-scalar ratio $r=0.1$ on the right panel.  }
\label{fig:cmb}
\end{figure}

In order to explicitly illustrate the effect of the ERE on the CMB power spectra, we keep $\tau_{\textrm{re}}+\Delta \tau=0.089$ fixed, and consider three different ionization histories:
$\Delta x_\textrm{e}(z)=0$ and $z_{\textrm{re}}=11$; $z_\text{re}=8$ and $z_\text{ere}=200$; $z_\text{re}=8$ and $z_\text{ere}=500$, where
\m
\Delta x_\textrm{e}(z)\sim \frac{1}{2}\left[\tanh\left(\frac{z_{\textrm{ere}}-z}{10}\right)+1\right]
\n
which are showed in Fig.~\ref{fig:xetanh}.
The CMB power spectra without tensor perturbations for these three ionization histories are showed on the left panel of Fig.~\ref{fig:cmb}. Similar to the scalar perturbations, we can also illustrate the contributions from the tensor perturbations to the CMB power spectra on the right panel of Fig.~\ref{fig:cmb}.
First of all, the temperature power spectra and the polarization power spectra at high $\ell$ for different ionization histories are almost the same respectively if $\tau_{\textrm{re}}+\Delta \tau$ is kept fixed. It is just what we expect. If the ERE happens, the E-mode polarization power spectra on intermediate scales are enhanced by $\sim \Delta\tau/(\eta_0-\eta_\text{ere})^2$. Since $\tau_{\textrm{re}}+\Delta \tau$ is kept fixed, the enhancements of the polarization power spectra on large scales $(\ell \lesssim 10)$ become smaller compared to that without the ERE. Finally, we need to mention that the ERE does not significantly enhance the CMB B-mode power spectrum on the intermediate scales if $\Delta \tau$ is not too large. 


\section{A Model-independent Analysis of the Early Reionization}
\label{pca}

In this section we will introduce the principal components analysis (PCA) and use this model-independent method to reconstruct the early reionization history.

We suppose that the ionization fraction due to the ERE takes the form
\begin{eqnarray}
\Delta x_e(z)=\sum_{i=1}^N\alpha_i\frac{1}{2}\left[\tanh\left(\frac{z_{i}-z}{\Delta z}\right)+1\right],
\label{eq:before}
\end{eqnarray}
where $\Delta z$ is the spacing between the $\{z_i\}$, and $\{\alpha_i\}$ are the coefficients. Here we assume that the ERE may happen in the range of $10<z<910$, and take $N=89$ and $\Delta z=10$. It implies that there are nighty redshift-bins covering this redshift range,  $z_1=20$ and $z_{89}=900$.

\begin{figure}[!htb]
\begin{center}
\includegraphics[width=9cm]{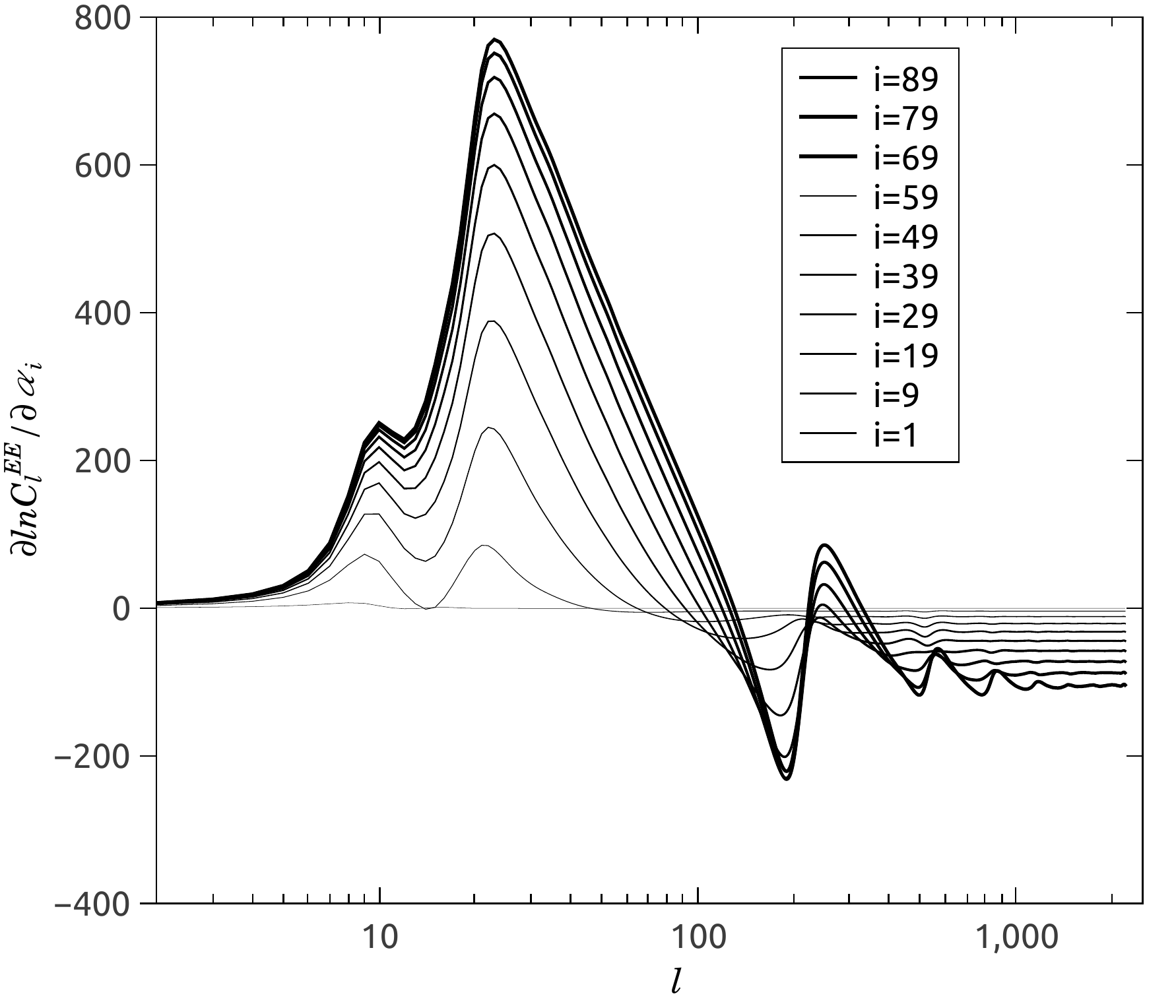}
\end{center}
\caption{The matrix of ${\partial \ln C_{\ell}^{EE}}/{\partial\alpha_i}$.}
\label{fig:transfer}
\end{figure}

Adopting Eq.~(\ref{eq:before}), we can compute the effect of nonzero $\{\alpha_i\}$ on the anisotropy spectrum by ${\partial \ln C_{\ell}}/{\partial\alpha_i}$ under a fiducial model in which the cosmological parameters are listed in Tab.~\ref{fiducial}.
\begin{table*}[!htp]
\centering
\renewcommand{\arraystretch}{1.5}
\begin{tabular}{ccccccc}
\hline\hline
$\Omega_bh^2$     &$\Omega_ch^2$   &$\Omega_\nu h^2$     &$H_0$     &$\tau_{re}$    &$n_s$     &$10^9A_s$\\
\hline
$0.02225$         &$0.1198$        &$0.00065$            &$67.27$   &$0.079$        &$0.9645$  &$2.207$ \\
\hline
\end{tabular}
\caption{The cosmological parameters in the fiducial model. }
\label{fiducial}
\end{table*}
According to the discussion in the former section, we notice that the E-mode polarization power spectrum is sensitive to the ERE, and therefore we focus on the matrix of ${\partial \ln C_{\ell}^{EE}}/{\partial\alpha_i}$ which are showed in Fig.~\ref{fig:transfer}.

\begin{figure}[!htb]
\begin{center}
\includegraphics[width=7cm]{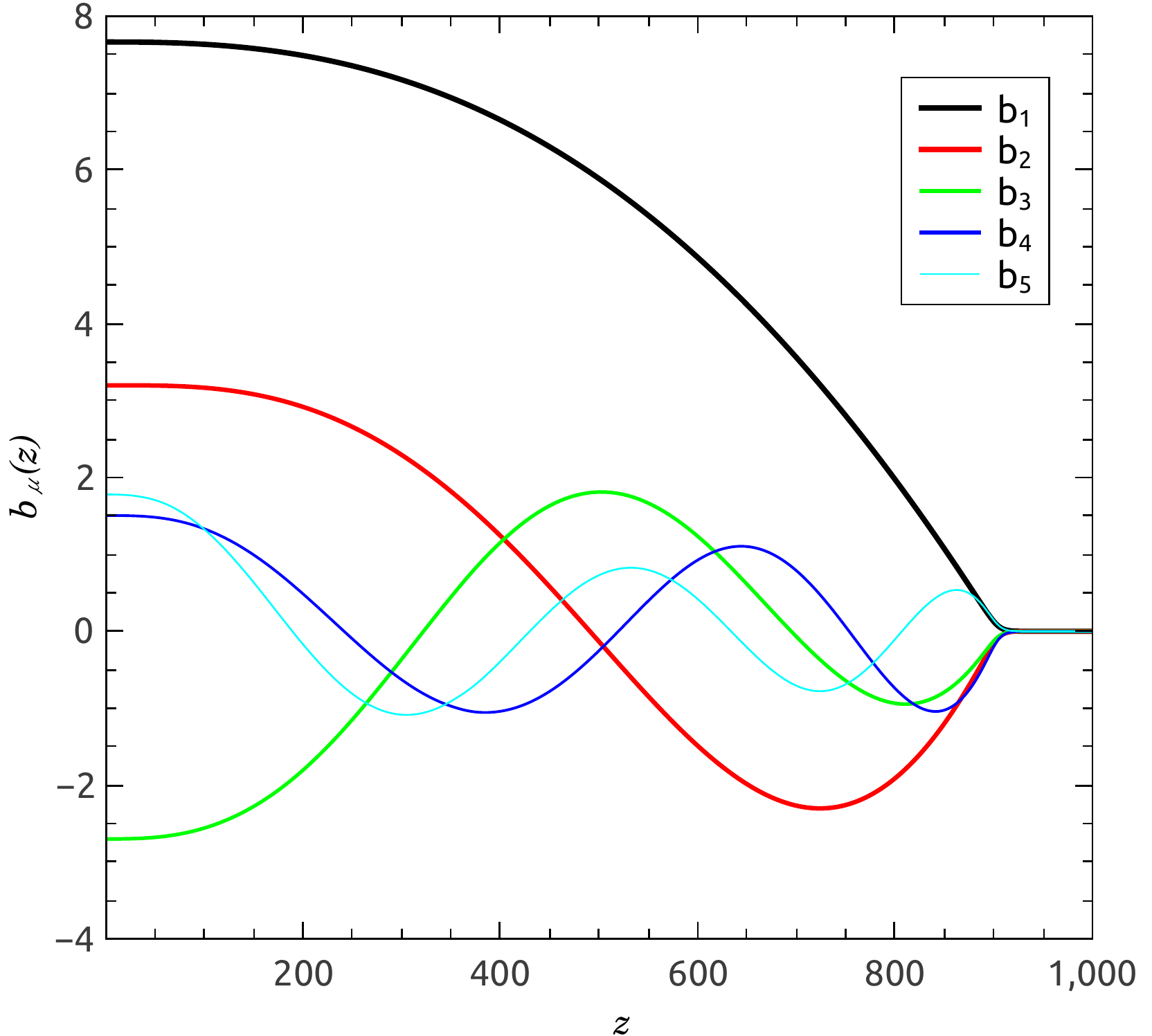}\quad \quad \includegraphics[width=6.7cm]{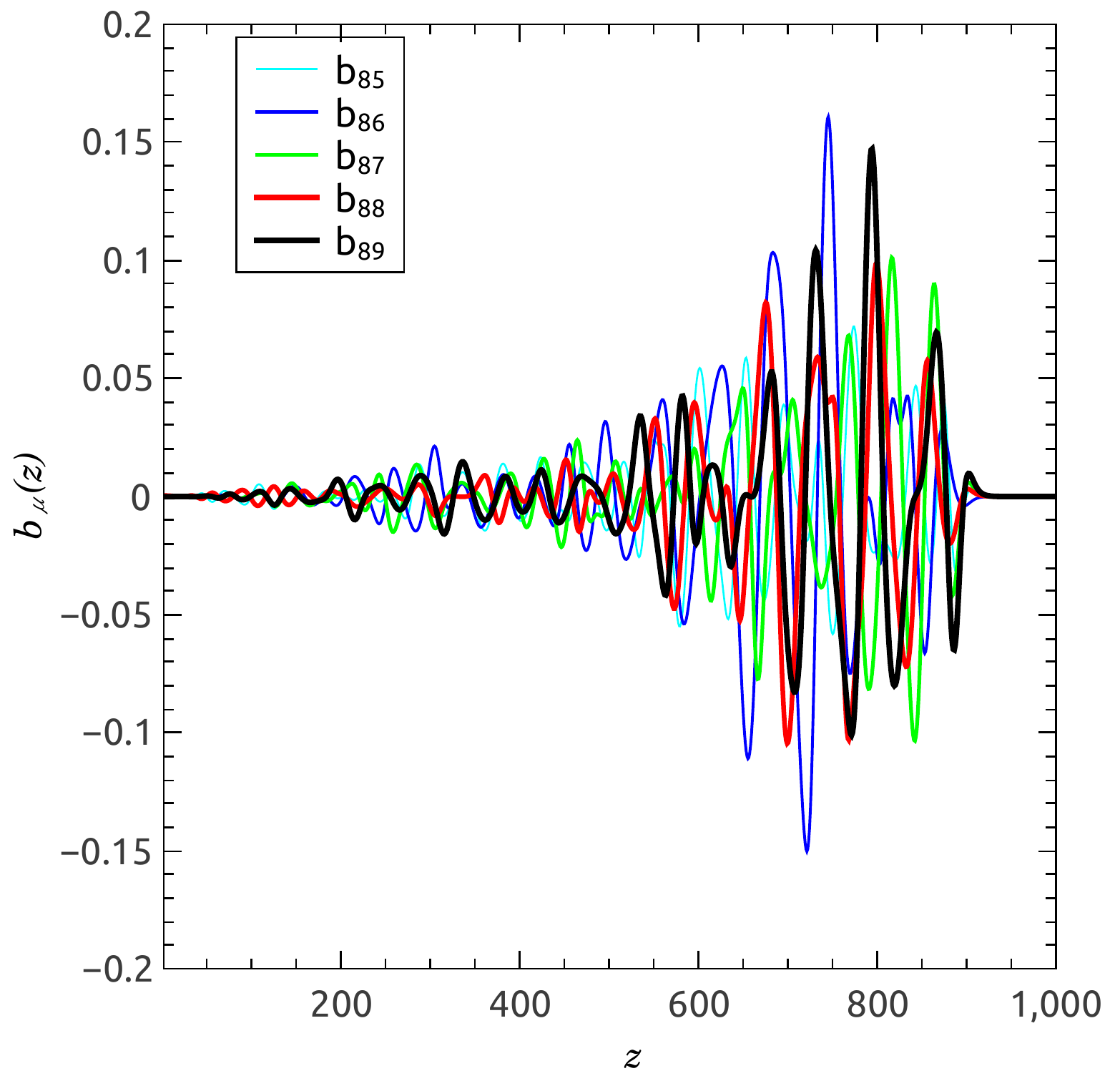}
\end{center}
\caption{The fist five new basis on the left panel and the last five new basis on the right panel.}
\label{fig:best}
\end{figure}

In principle, in turn, we should use E-mode polarization power spectrum to estimate the $\{\alpha_i\}$. However, in practice, the large number of free parameters including usual CMB parameters and $\{\alpha_i\}$ make parameter estimation impossible in a likelihood analysis. Fortunately, we can turn to the Fisher matrix of all-sky polarization experiment for the ERE
\begin{eqnarray}
F_{ij}=\sum_{\ell}({\ell}+\frac{1}{2})\frac{\partial \ln C_{\ell}^{EE}}{\partial\alpha_i}\frac{\partial \ln C_{\ell}^{EE}}{\partial\alpha_j}.
\label{eq:fisher}
\end{eqnarray}
Diagonalizing the matrix $F$ by an orthogonal matrix $S$
\begin{eqnarray}
\Lambda=S^TFS,
\end{eqnarray}
the old basis of $\left\{\frac{1}{2}\left[\tanh\left(\frac{z_{i}-z}{\Delta z}\right)+1\right]\right\}$ and the new basis of $\{b_\mu(z)\}$ can be related to $S$ by
\begin{eqnarray}
b_\mu(z)=\sum_i^N{\frac{1}{2}\left[\tanh\left(\frac{z_{i}-z}{\Delta z}\right)+1\right]}S_{i\mu},
\end{eqnarray}
and $\Delta x_e(z)$ can be represented by the new basis of $\{b_\mu\}$ as
\begin{eqnarray}
\label{eq:representation2}
\Delta x_e(z)=\sum_\mu^{N_\beta} \beta_\mu{b_\mu}(z).
\end{eqnarray}
Since the $i$-th column of $S$ is the eigenvector of $F$ corresponding to the eigenvalue $\lambda_i$, we can fix $S$ by ordering $\{\lambda_i\}$ to be $\lambda_1>\lambda_2>...>\lambda_{N_\beta}$. The largest eigenvalues contain the most information of the ERE. The fist five and last five new basis are illustrated in Fig.~\ref{fig:best}.

\begin{figure}[!htb]
\begin{center}
\includegraphics[width=6cm]{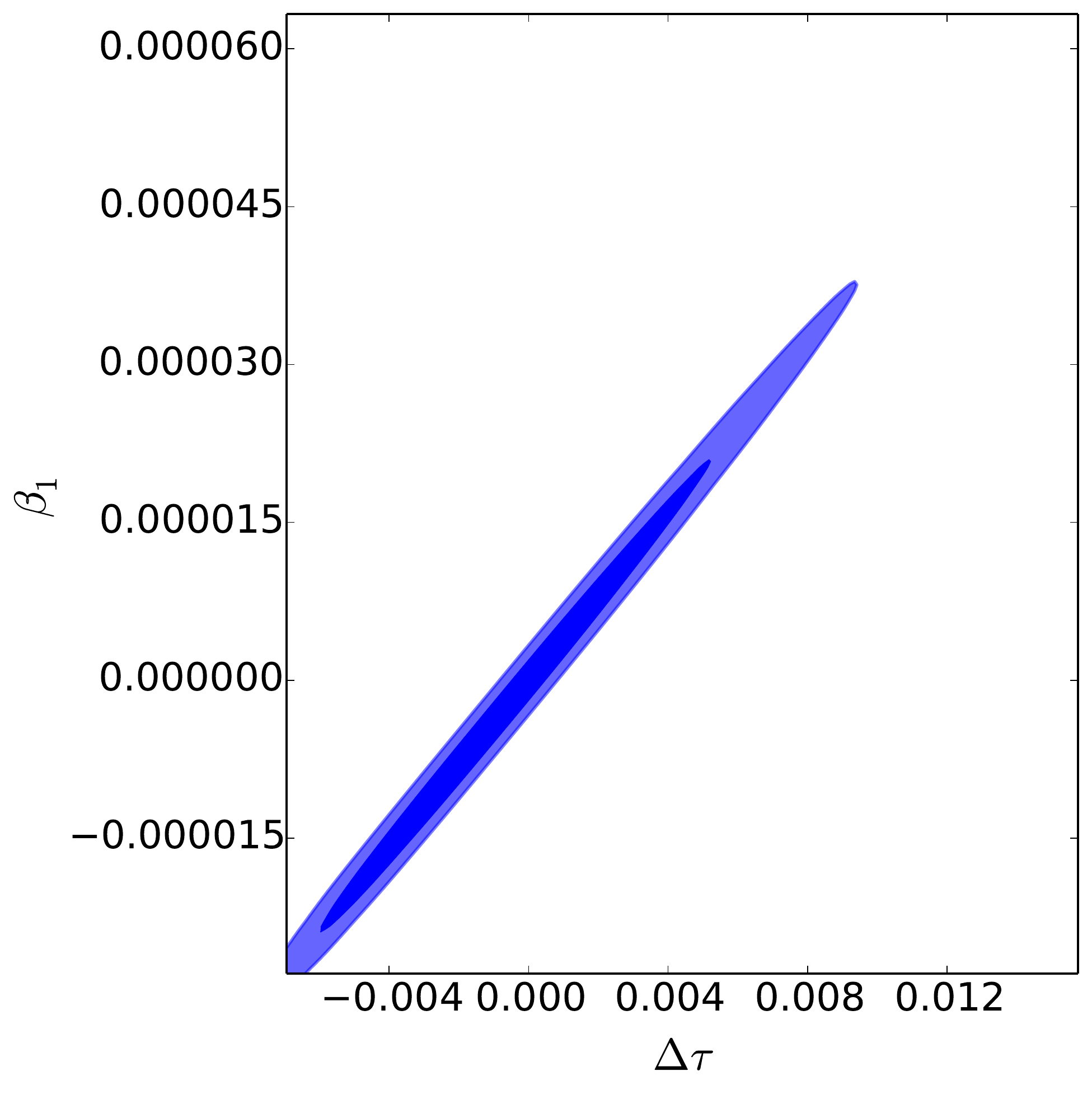}
\end{center}
\caption{Constraints on the coefficient of the first basis and $\Delta\tau$ in $_{\beta_1}\Lambda$CDM model from $\textit{Planck}$ TT,TE,EE+lowP.}
\label{fig:fit1}
\end{figure}
\begin{figure}[!htb]
\begin{center}
\includegraphics[width=10cm]{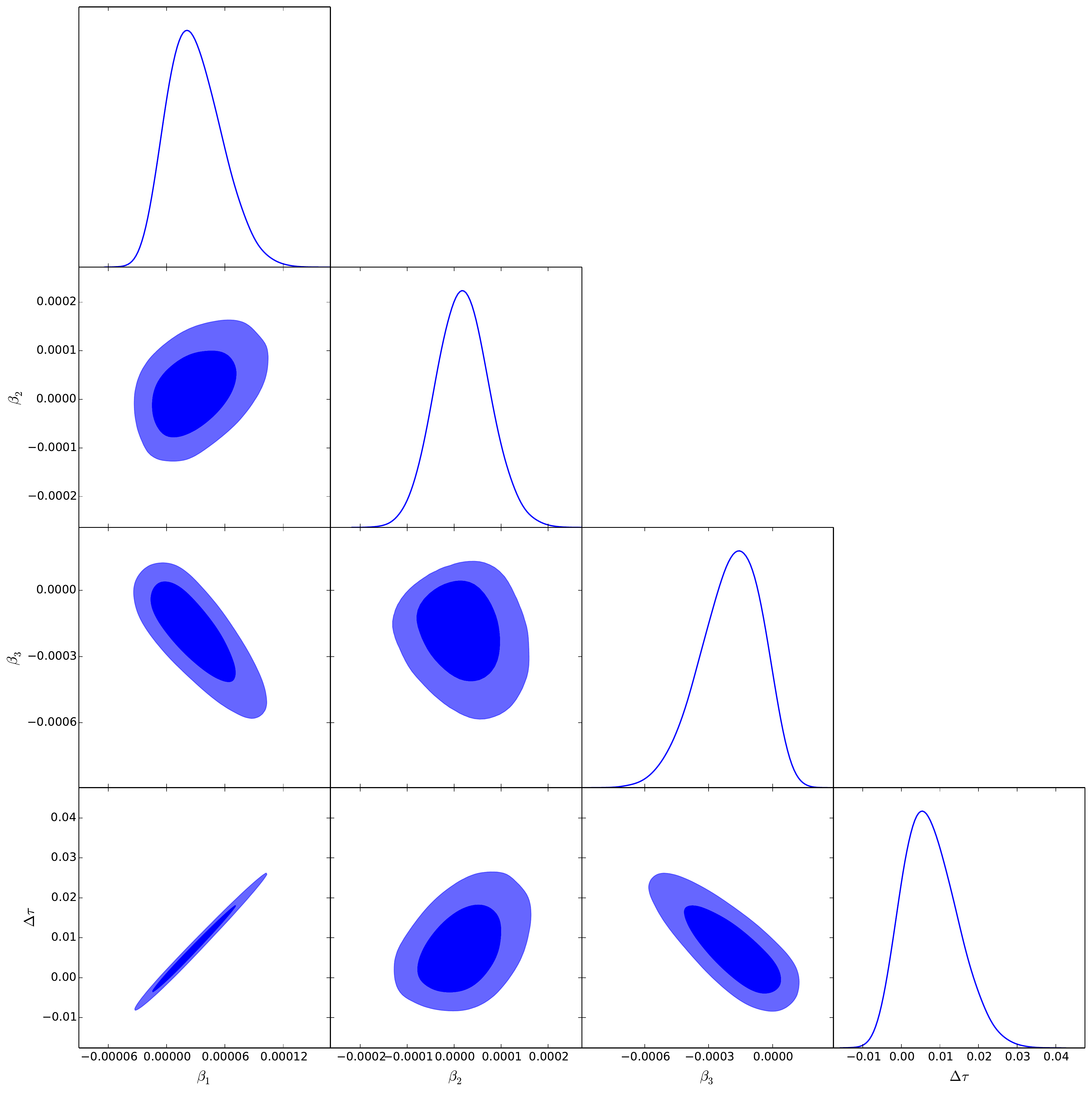}
\end{center}
\caption{Constraints on the coefficients of the first three basis and $\Delta\tau$ in $_{\beta_3}\Lambda$CDM model from $\textit{Planck}$ TT,TE,EE+lowP.}
\label{fig:fit3}
\end{figure}
\begin{figure}[!htb]
\begin{center}
\includegraphics[width=12cm]{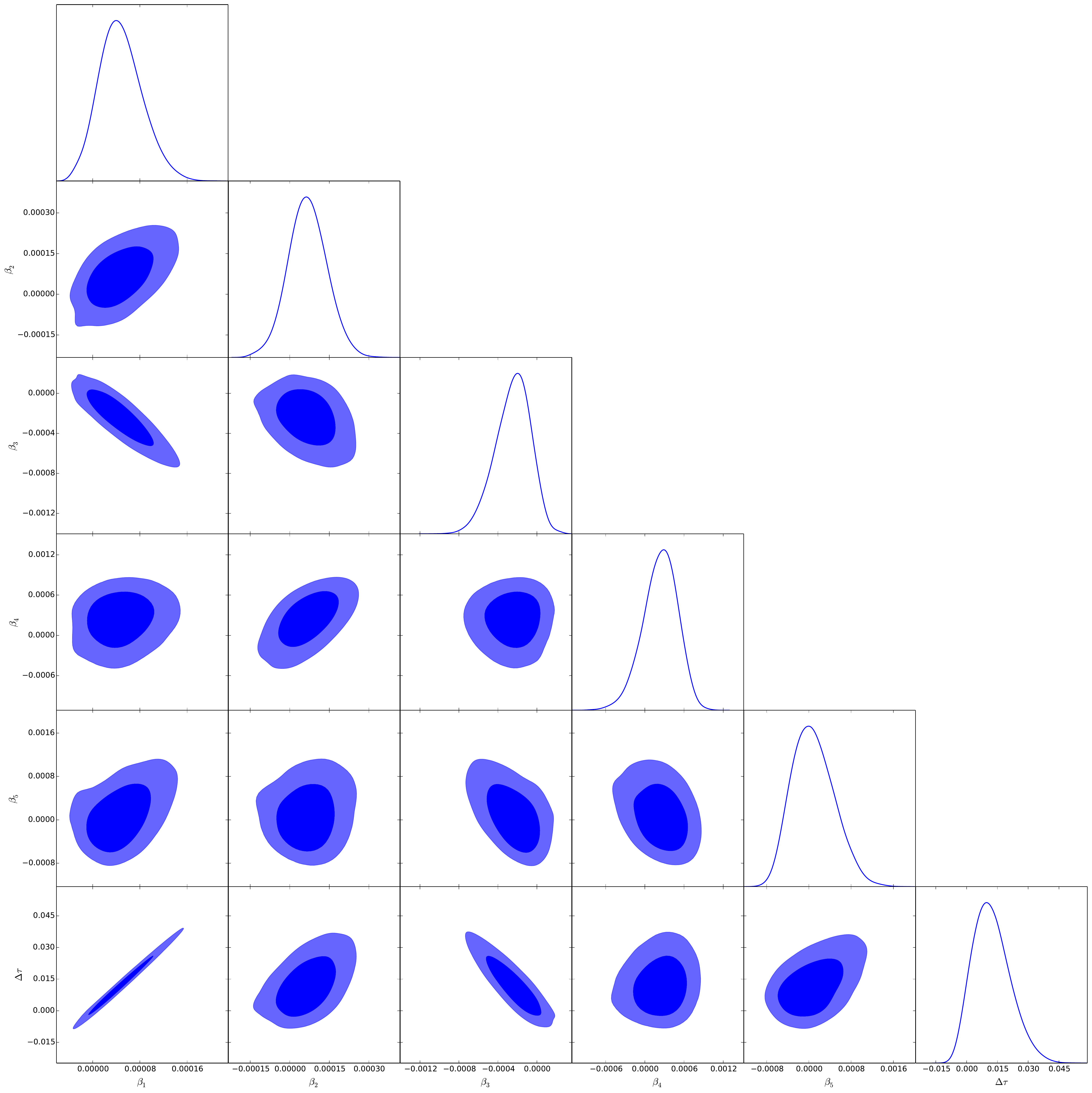}
\end{center}
\caption{Constraints on the coefficients of the first five basis and $\Delta\tau$ in $_{\beta_5}\Lambda$CDM model from $\textit{Planck}$ TT,TE,EE+lowP.}
\label{fig:fit5}
\end{figure}
\begin{figure}[!htb]
\begin{center}
\includegraphics[width=9cm]{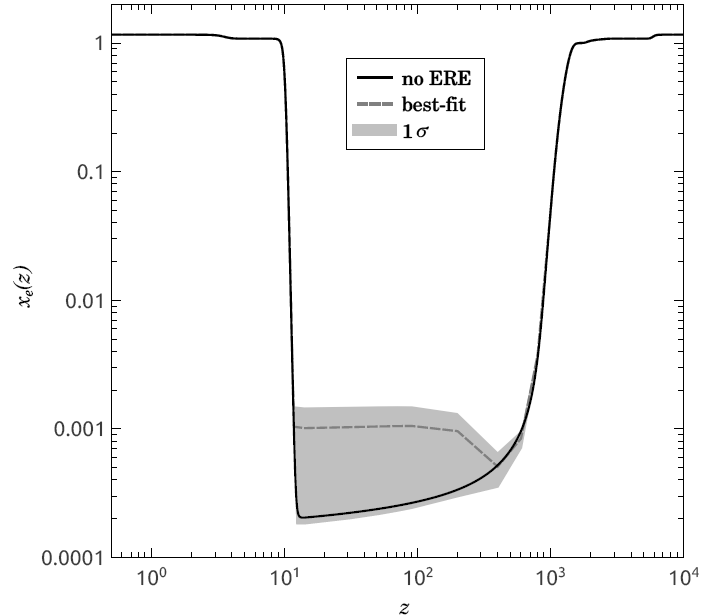}
\end{center}
\caption{Ionization history of universe in $_{\beta_3}\Lambda$CDM model.}
\label{fig:erexe}
\end{figure}

Now we are ready to reconstruct the ionization history during the ERE. Utilizing the new parameterization of $\Delta x_e(z)$ in Eq.~(\ref{eq:representation2}), we can use $\textit{Planck}$ TT,TE,EE+lowP released in 2015 \cite{Ade:2015xua} to constrain the coefficients $\{\beta_\mu\}$ of the basis $\{b_\mu\}$. Here we consider three cases: $N_\beta=1$ denoted by $_{\beta_1}\Lambda$CDM model; $N_\beta=3$ denoted by $_{\beta_3}\Lambda$CDM model; $N_\beta=5$ denoted by $_{\beta_5}\Lambda$CDM model. Note that the values of $\{\beta_\mu\}$ in every model must satisfy the condition of $0\leq x_e(z)\lesssim1.16$. Our results are included in Tab.~\ref{tab:PC} and Figs.~\ref{fig:fit1}, \ref{fig:fit3} and \ref{fig:fit5}.
\begin{table*}[!htp]
\centering
\renewcommand{\arraystretch}{1.5}
\scalebox{1}[1]{%
\begin{tabular}{|c|c c c|}
\hline
                                       &$_{\beta_1}\Lambda$CDM        &$_{\beta_3}\Lambda$CDM          &$_{\beta_5}\Lambda$CDM      \\
\hline
$\Omega_bh^2$                          &$0.02224\pm0.00016$        &$0.02223\pm0.00016$          &$0.02223\pm0.00016$      \\
$\Omega_ch^2$                          &$0.1199\pm0.0016$          &$0.1202\pm0.0016$            &$0.1202\pm0.0016$        \\
$100\theta_{\emph{MC}}$                &$1.04076\pm0.00033$        &$1.04065\pm0.00035$          &$1.04070\pm0.00038$      \\
$z_{re}$                               &$9.98\pm1.58$              &$9.55\pm1.66$                &$9.17\pm1.66$            \\
${\textrm{ln}}(10^{10}A_s)$            &$3.092\pm0.033$            &$3.098\pm0.033$              &$3.100\pm0.033$          \\
$n_s$                                  &$0.9643\pm0.0047$          &$0.9653\pm0.0050$            &$0.9642\pm0.0052$        \\
$\beta_1$                              &$0.000000\pm0.000014$      &$0.000030\pm0.000027$        &$0.000048\pm0.000036$    \\
$\beta_2$                              &-                          &$0.000018\pm0.000056$        &$0.000068\pm0.000072$    \\
$\beta_3$                              &-                          &-$0.000179\pm0.000141$       &-$0.000249\pm0.000179$   \\
$\beta_4$                              &-                          &-                            &$0.000232\pm0.000262$    \\
$\beta_5$                              &-                          &-                            &$0.000087\pm0.000390$    \\
\hline
$\Delta\tau$                           &$0.000\pm0.004$            &$0.008\pm0.007$              &$0.012\pm0.009$          \\
$\tau_{\textrm{re}}+\Delta\tau$        &$0.079\pm0.017$            &$0.081\pm0.017$              &$0.082\pm0.017$          \\
\hline
\end{tabular}}
\caption{The $68\%$ limits on the cosmological parameters in $_{\beta_1}\Lambda$CDM model, $_{\beta_3}\Lambda$CDM model and $_{\beta_5}\Lambda$CDM model from $\textit{Planck}$ TT,TE,EE+lowP datasets.}
\label{tab:PC}
\end{table*}
From these results, there is no evidence for the ERE and the instantaneous reionization is quite consistent with the data. At $95\%$ confidence level, the limits on $\Delta \tau$ are $\Delta \tau<0.007$ for $_{\beta_1}\Lambda$CDM model, $\Delta \tau<0.022$ for $_{\beta_3}\Lambda$CDM model, and $\Delta \tau<0.031$ for $_{\beta_5}\Lambda$CDM model, respectively. Finally, for an instance, the ionization history for $_{\beta_3}\Lambda$CDM model are illustrated in Fig.~\ref{fig:erexe}.
From Fig.~~\ref{fig:erexe}, there is still a room for the ERE in the range of $10<z\lesssim 500$.



\section{Effect of the Early Reionization on the Cosmological Parameter Estimates}
\label{fitting}

In this section we investigate how the ERE affects the cosmological parameter estimates. Because the ERE mainly disturbs the CMB polarization power spectra on the intermediate and large scales, we take the ERE into account and explore how it will modify the constraints on the tensor-to-scalar ratio $r$ and the neutrino mass respectively. Our results are summarized in Tab.~\ref{tab:rnu}.
\begin{table*}[!htp]
\centering
\renewcommand{\arraystretch}{1.5}
\scalebox{1}[1]{%
\begin{tabular}{|c|c|c|c|c|}
\hline
&\multicolumn{2}{c|}{\tiny $\textit{Planck}$ TT,TE,EE+lowP+lensing, BICEP2\&\textit{Keck Array} and BAO}
&\multicolumn{2}{c|}{\tiny $\textit{Planck}$ TT,TE,EE+lowP and BAO}                                                  \\
\cline{2-5}
                                &$\Lambda$CDM$+r$      &$_{\beta_3}\Lambda$CDM$+r$ &$\Lambda$CDM$+\sum m_\nu$ &$_{\beta_3}\Lambda$CDM$+\sum m_\nu$      \\
\hline
$\Omega_bh^2$                   &$0.02231\pm0.00014$   &$0.02231\pm0.00014$        &$0.02233\pm0.00014$       &$0.02232\pm0.00015$\\
$\Omega_ch^2$                   &$0.1184\pm0.0010$     &$0.1182\pm0.0011$          &$0.1187\pm0.0011$         &$0.1187\pm0.0012$\\
$100\theta_{\emph{MC}}$         &$1.04096\pm0.00030$   &$1.04095\pm0.00031$        &$1.04091\pm0.00030$       &$1.04085\pm0.00032$\\
$z_{re}$                        &$9.14\pm1.11$         &$8.98\pm1.26$              &$10.47\pm1.47$            &$10.14\pm1.56$ \\
${\textrm{ln}}(10^{10}A_s)$     &$3.122\pm0.021$       &$3.123\pm0.021$            &$3.102\pm0.033$           &$3.104\pm0.033$ \\
$n_s$                           &$0.9681\pm0.0040$     &$0.9690\pm0.0045$          &$0.9676\pm0.0042$         &$0.9683\pm0.0045$ \\
$\beta_1$                       &-                     &$0.000013\pm0.000024$      &-                         &$0.000020\pm0.000027$ \\
$\beta_2$                       &-                     &$0.000026\pm0.000053$      &-                         &$0.000022\pm0.000058$ \\
$\beta_3$                       &-                     &-$0.000150\pm0.000125$     &-                         &-$0.000173\pm0.000133$\\
$r_{0.01}~(95\%)$                      &$<0.071$             &$<0.074$                  &-                         &-\\
$\sum m_\nu~(95\%)$             &-                     &-                          &$<0.140~\textrm{eV}$      &$<0.139~\textrm{eV}$\\
\hline
$\Delta\tau$                    &-                   &$0.003\pm0.006$            &-                       &$0.005\pm0.007$\\
$\tau_{\textrm{re}}+\Delta\tau$ &$0.070\pm0.012$       &$0.072\pm0.012$            &$0.084\pm0.017$           &$0.086\pm0.017$\\
\hline
\end{tabular}}
\caption{The $68\%$ (or $95\%$) limits on the cosmological parameters in $\Lambda$CDM$+r$ model and $_{\beta_3}\Lambda$CDM$+r$ model from the data combination of $\textit{Planck}$ TT,TE,EE+lowP+lensing, BICEP2\&\textit{Keck Array} and BAO. And the $68\%$ (or $95\%$) limits on the cosmological parameters in $\Lambda$CDM$+\sum m_\nu$ model and $_{\beta_3}\Lambda$CDM$+\sum m_\nu$ model from the data combination of $\textit{Planck}$ TT,TE,EE+lowP and BAO.}
\label{tab:rnu}
\end{table*}

\begin{figure}[!htb]
\begin{center}
\includegraphics[width=9cm]{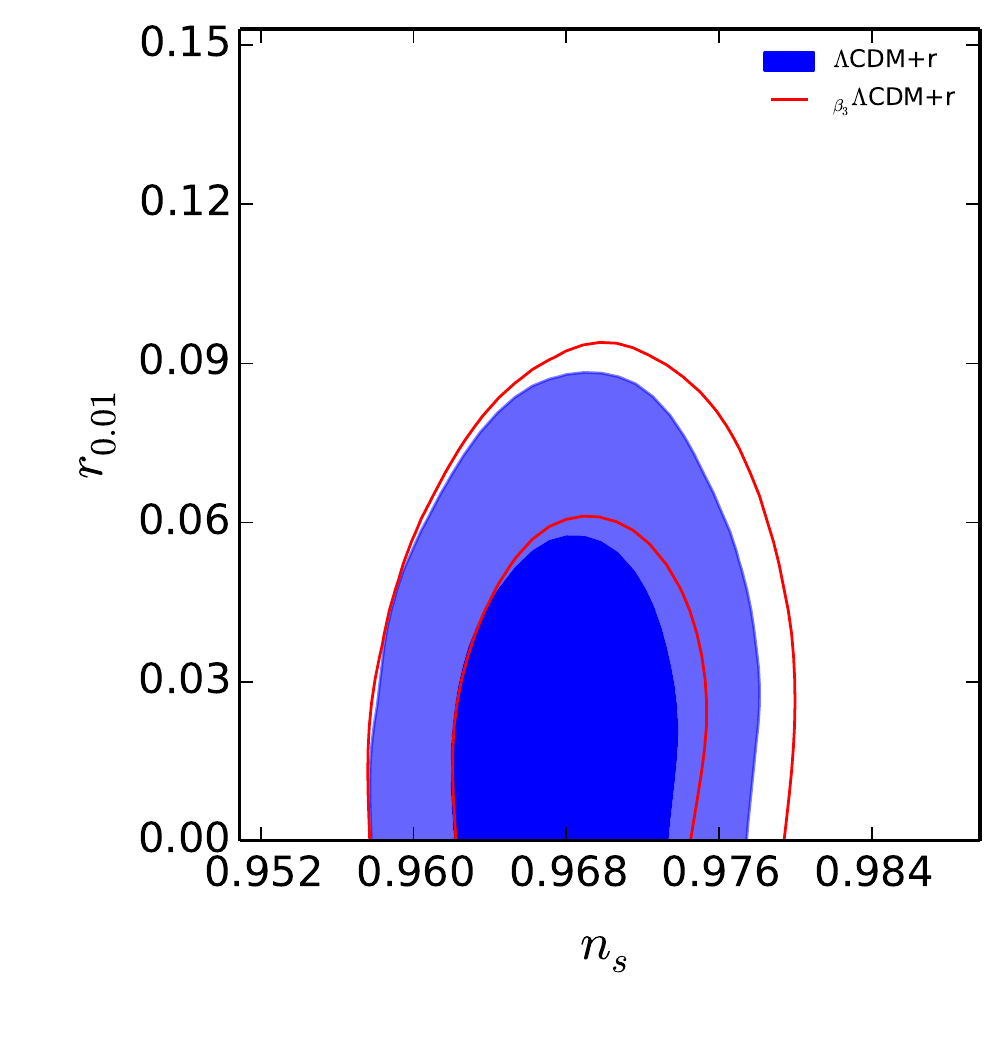}
\end{center}
\caption{Constraints on the cosmological parameters in $\Lambda$CDM$+r$ model and $_{\beta_3}\Lambda$CDM$+r$ model from the combination of Planck, BICEP2 \& \textit{Keck Array} and BAO datasets.}
\label{fig:fit3r}
\end{figure}

Primordial gravitational waves can be generated during inflation in the very early universe, and the amplitude of gravitational-wave power spectrum is parametrized by the so-called tensor-to-scalar ratio $r$. The primordial gravitational waves can contribute to the CMB B-modes mainly on the intermediate and large scales. In the $\Lambda$CDM+$r$ model with instantaneous reionization at low redshift and the pivot scale $k_p=0.01\textrm{Mpc}^{-1}$, the constraint on $r$ is
\m
r_{0.01}<0.071
\n
at $95\%$ confidence level (CL) by combining $\textit{Planck}$ TT,TE,EE+lowP+lensing, BICEP2 \& \textit{Keck Array} \cite{Array:2015xqh} and BAO including 6dFGS \cite{Beutler:2011hx}, MGS \cite{Ross:2014qpa}, LOWZ and CMASS of BOSS DR12 \cite{Cuesta:2015mqa,Gil-Marin:2015nqa}.
Furthermore, we consider an extended cosmological models, namely $_{\beta_3}\Lambda$CDM$+r$ model, and see how the ERE affects the constraint on the tensor-to-scalar ratio $r$. The results is
\m
r_{0.01}<0.074
\n
at $95\%$CL in $_{\beta_3}\Lambda$CDM$+r$ model. The contour plots show up in Fig.~\ref{fig:fit3r}.
We find that the constraints on $r$ and $n_s$ in the model with ERE do not significantly change compared to those in the model with the instantaneous reionization.

\begin{figure}[!htb]
\begin{center}
\includegraphics[width=8cm]{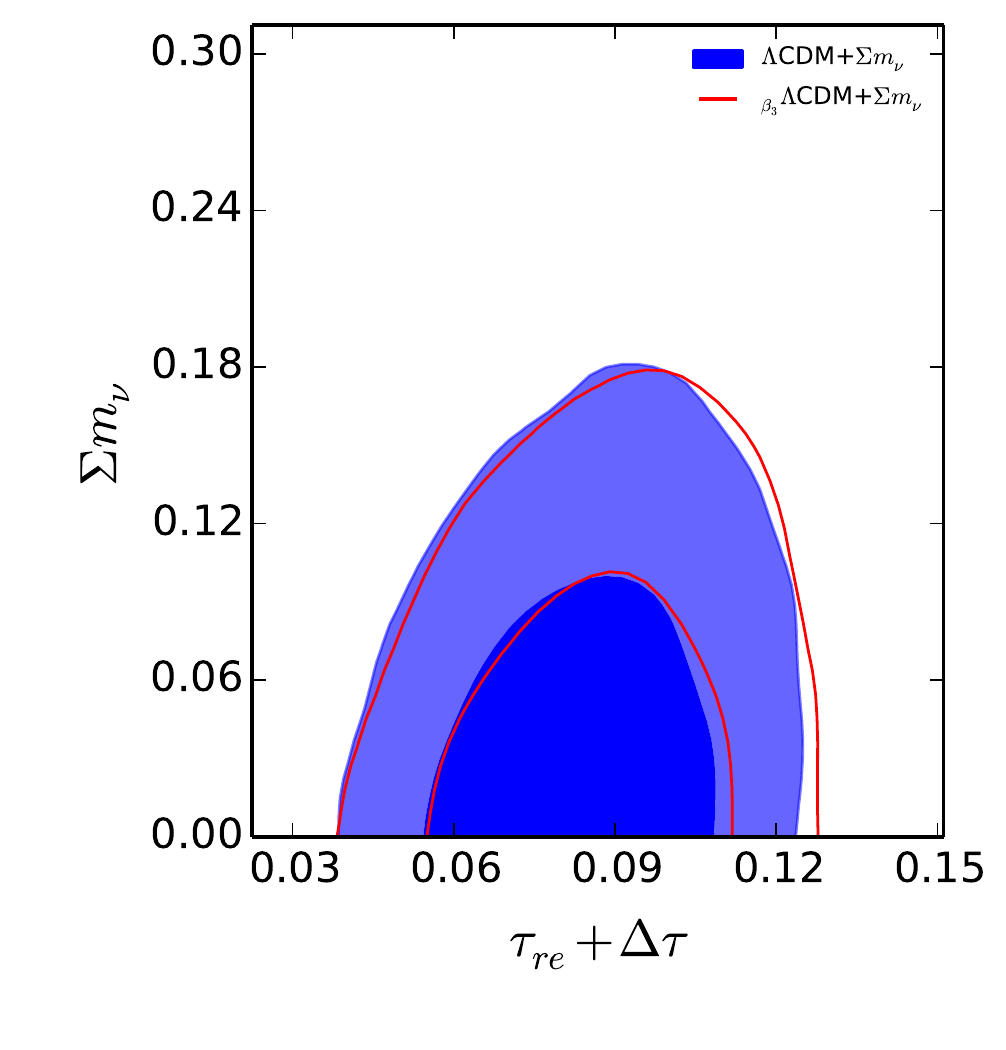}
\end{center}
\caption{Constraints on the neutrino mass in $\Lambda$CDM$+\sum m_\nu$ model and $_{\beta_3}\Lambda$CDM$+\sum m_\nu$ model.}
\label{fig:fit3nu}
\end{figure}

The main signature of massive neutrinos in the CMB comes about via the early ISW effect, and it is worthy considering how the ERE affects the constraint on the neutrino mass. Again we constrain the neutrino mass in the instantaneous reionization model and find
\begin{equation}
\sum m_\nu<0.140~\textrm{eV}
\end{equation}
at $95\%$ CL by combining $\textit{Planck}$ TT,TE,EE+lowP and BAO datasets. The constraints become
\begin{equation}
\sum m_\nu<0.139~\textrm{eV}
\end{equation}
at $95\%$ CL in $_{\beta_3}\Lambda$CDM$+\sum m_\nu$ model. See the contour plots in Fig.~\ref{fig:fit3nu} and constraints on the free parameters in Tab.~\ref{tab:rnu}.
We see that the constraint on the neutrino mass becomes just slightly tighter in $_{\beta_3}\Lambda$CDM$+\sum m_\nu$ model than that in the instantaneous reionization model.

\section{Summary and discussion}
\label{sd}

Even though the instantaneous reionization at redshift less than ten is consistent with the data, the possibility of a reionization which occurs at higher redshifts but after recombination due to the accretion of gas onto primordial black holes and/or annihilation of dark matter etc is still allowed. In this paper we find that the ERE mainly disturbs the CMB polarization power spectra on the intermediate and large scales if the total optical depth is kept fixed. Adopting the Planck polarization data, we model-independently reconstruct the ionization history during the ERE, and find that an order of $10^{-2}$ optical depth due to the ERE is still allowed.

In addition, we also explore how the ERE affects the cosmological parameter estimates. Because both the tensor perturbations and the neutrino mass disturb the CMB power spectra on the intermediate and large scales, we take the ERE into account and constrain the tensor-to-scalar ratio and the neutrino mass by adopting the currently available cosmological data. We find that the ERE does not significantly change the constraints on cosmological parameters at the sensitivities of current experiments. However, a tighter constraint on the ERE might be important if we want to more precisely constrain the cosmological parameters in the future.

\vspace{0.5 cm}
\noindent {\bf Acknowledgments}

We acknowledge the use of HPC Cluster of SKLTP/ITP-CAS.
This work is supported by grants from NSFC (grant NO. 11335012, 11575271, 11690021), Top-Notch Young Talents Program of China, and partly supported by Key Research Program of Frontier Sciences, CAS.

\newpage




\begin{thebibliography}{99}
\frenchspacing

\bibitem{Fan:2005es}
  X.~H.~Fan {\it et al.},
  Astron.\ J.\  {\bf 132}, 117 (2006)  doi:10.1086/504836  [astro-ph/0512082].  

\bibitem{McGreer:2014qwa}
  I.~McGreer, A.~Mesinger and V.~D'Odorico,
  Mon.\ Not.\ Roy.\ Astron.\ Soc.\  {\bf 447}, no. 1, 499 (2015)  doi:10.1093/mnras/stu2449  [arXiv:1411.5375 [astro-ph.CO]].  

\bibitem{Schroeder:2012uy}
  J.~Schroeder, A.~Mesinger and Z.~Haiman,
  Mon.\ Not.\ Roy.\ Astron.\ Soc.\  {\bf 428}, 3058 (2013)  doi:10.1093/mnras/sts253  [arXiv:1204.2838 [astro-ph.CO]].  

\bibitem{Pentericci:2014nia}
  L.~Pentericci {\it et al.},
  Astrophys.\ J.\  {\bf 793}, no. 2, 113 (2014)  doi:10.1088/0004-637X/793/2/113  [arXiv:1403.5466 [astro-ph.CO]].  

\bibitem{Schenker:2014tda}
  M.~A.~Schenker, R.~S.~Ellis, N.~P.~Konidaris and D.~P.~Stark,
  Astrophys.\ J.\  {\bf 795}, no. 1, 20 (2014)  doi:10.1088/0004-637X/795/1/20  [arXiv:1404.4632 [astro-ph.CO]].  

\bibitem{Tilvi:2014oia}
  V.~Tilvi {\it et al.},
  Astrophys.\ J.\  {\bf 794}, no. 1, 5 (2014)  doi:10.1088/0004-637X/794/1/5  [arXiv:1405.4869 [astro-ph.CO]].  

\bibitem{Robertson:2013bq}
  B.~E.~Robertson {\it et al.},
  Astrophys.\ J.\  {\bf 768}, 71 (2013)  doi:10.1088/0004-637X/768/1/71  [arXiv:1301.1228 [astro-ph.CO]].  

\bibitem{Robertson:2015uda}
  B.~E.~Robertson, R.~S.~Ellis, S.~R.~Furlanetto and J.~S.~Dunlop,
  Astrophys.\ J.\  {\bf 802}, no. 2, L19 (2015)  doi:10.1088/2041-8205/802/2/L19  [arXiv:1502.02024 [astro-ph.CO]].  

\bibitem{Ade:2015xua}
  P.~A.~R.~Ade {\it et al.} [Planck Collaboration],
  Astron.\ Astrophys.\  {\bf 594}, A13 (2016)  doi:10.1051/0004-6361/201525830  [arXiv:1502.01589 [astro-ph.CO]].  

\bibitem{Heinrich:2016ojb}
  C.~H.~Heinrich, V.~Miranda and W.~Hu,
  arXiv:1609.04788 [astro-ph.CO].  

\bibitem{Miranda:2016trf}
  V.~Miranda, A.~Lidz, C.~H.~Heinrich and W.~Hu,
  arXiv:1610.00691 [astro-ph.CO].  

\bibitem{Slatyer:2009yq}
  T.~R.~Slatyer, N.~Padmanabhan and D.~P.~Finkbeiner,

\bibitem{Chluba:2009uv}
  J.~Chluba,
  Mon.\ Not.\ Roy.\ Astron.\ Soc.\  {\bf 402}, 1195 (2010)  doi:10.1111/j.1365-2966.2009.15957.x  [arXiv:0910.3663 [astro-ph.CO]].  

\bibitem{Finkbeiner:2011dx}
  D.~P.~Finkbeiner, S.~Galli, T.~Lin and T.~R.~Slatyer,
  Phys.\ Rev.\ D {\bf 85}, 043522 (2012)  doi:10.1103/PhysRevD.85.043522  [arXiv:1109.6322 [astro-ph.CO]].  

\bibitem{Liu:2016cnk}
  H.~Liu, T.~R.~Slatyer and J.~Zavala,
  Phys.\ Rev.\ D {\bf 94}, no. 6, 063507 (2016)  doi:10.1103/PhysRevD.94.063507  [arXiv:1604.02457 [astro-ph.CO]].  

\bibitem{Oldengott:2016yjc}
  I.~M.~Oldengott, D.~Boriero and D.~J.~Schwarz,
  JCAP {\bf 1608}, no. 08, 054 (2016)  doi:10.1088/1475-7516/2016/08/054  [arXiv:1605.03928 [astro-ph.CO]].  

\bibitem{Ricotti:2007au}
  M.~Ricotti, J.~P.~Ostriker and K.~J.~Mack,
  Astrophys.\ J.\  {\bf 680}, 829 (2008)  doi:10.1086/587831  [arXiv:0709.0524 [astro-ph]].  

\bibitem{Chen:2016pud}
  L.~Chen, Q.~G.~Huang and K.~Wang,
  JCAP {\bf 1612}, no. 12, 044 (2016)
  doi:10.1088/1475-7516/2016/12/044
  [arXiv:1608.02174 [astro-ph.CO]].

\bibitem{Ali-Haimoud:2016mbv}
  Y.~Ali-Haimoud and M.~Kamionkowski,
  Phys.\ Rev.\ D {\bf 95}, no. 4, 043534 (2017)
  doi:10.1103/PhysRevD.95.043534
  [arXiv:1612.05644 [astro-ph.CO]].


\bibitem{Smith:2006nk}
  K.~M.~Smith, W.~Hu and M.~Kaplinghat,
  Phys.\ Rev.\ D {\bf 74}, 123002 (2006)  doi:10.1103/PhysRevD.74.123002  [astro-ph/0607315].  

\bibitem{Allison:2015qca}
  R.~Allison, P.~Caucal, E.~Calabrese, J.~Dunkley and T.~Louis,
  Phys.\ Rev.\ D {\bf 92}, no. 12, 123535 (2015)  doi:10.1103/PhysRevD.92.123535  [arXiv:1509.07471 [astro-ph.CO]].  

\bibitem{Kamionkowski:2015yta}
  M.~Kamionkowski and E.~D.~Kovetz,
  doi:10.1146/annurev-astro-081915-023433  arXiv:1510.06042 [astro-ph.CO].  

\bibitem{Mortonson:2009xk}
  M.~J.~Mortonson and W.~Hu,
  Phys.\ Rev.\ D {\bf 80}, 027301 (2009)  doi:10.1103/PhysRevD.80.027301  [arXiv:0906.3016 [astro-ph.CO]].  

\bibitem{Mortonson:2009qv}
  M.~J.~Mortonson, C.~Dvorkin, H.~V.~Peiris and W.~Hu,
  Phys.\ Rev.\ D {\bf 79}, 103519 (2009)  doi:10.1103/PhysRevD.79.103519  [arXiv:0903.4920 [astro-ph.CO]].  

\bibitem{Mortonson:2007tb}
  M.~J.~Mortonson and W.~Hu,
  Phys.\ Rev.\ D {\bf 77}, 043506 (2008)  doi:10.1103/PhysRevD.77.043506  [arXiv:0710.4162 [astro-ph]].  

\bibitem{Hu:2003gh}
  W.~Hu and G.~P.~Holder,
  Phys.\ Rev.\ D {\bf 68}, 023001 (2003)  doi:10.1103/PhysRevD.68.023001  [astro-ph/0303400].  

\bibitem{Dai:2015dwa}
  W.~M.~Dai, Z.~K.~Guo and R.~G.~Cai,
  Phys.\ Rev.\ D {\bf 92}, no. 12, 123521 (2015)  doi:10.1103/PhysRevD.92.123521  [arXiv:1509.01501 [astro-ph.CO]].  

\bibitem{Liu:2015gho}
  Y.~Liu, H.~Li, S.~Y.~Li, Y.~P.~Li and X.~Zhang,
  JCAP {\bf 1602}, no. 02, 046 (2016)  doi:10.1088/1475-7516/2016/02/046  [arXiv:1512.07394 [astro-ph.CO]].  

\bibitem{Lewis:2008wr}
  A.~Lewis,
  Phys.\ Rev.\ D {\bf 78}, 023002 (2008)
  doi:10.1103/PhysRevD.78.023002
  [arXiv:0804.3865 [astro-ph]].

\bibitem{Seager:1999bc}
  S.~Seager, D.~D.~Sasselov and D.~Scott,
  Astrophys.\ J.\  {\bf 523}, L1 (1999)  doi:10.1086/312250  [astro-ph/9909275].  

\bibitem{Seljak:1996is}
  U.~Seljak and M.~Zaldarriaga,
  Astrophys.\ J.\  {\bf 469}, 437 (1996)  doi:10.1086/177793  [astro-ph/9603033].  

\bibitem{Zaldarriaga:1996xe}
  M.~Zaldarriaga and U.~Seljak,
  Phys.\ Rev.\ D {\bf 55}, 1830 (1997)  doi:10.1103/PhysRevD.55.1830  [astro-ph/9609170].  

\bibitem{Array:2015xqh}
  P.~A.~R.~Ade {\it et al.} [BICEP2 and Keck Array Collaborations],
  Phys.\ Rev.\ Lett.\  {\bf 116}, 031302 (2016)
  doi:10.1103/PhysRevLett.116.031302
  [arXiv:1510.09217 [astro-ph.CO]].

\bibitem{Beutler:2011hx}
  F.~Beutler {\it et al.},
  Mon.\ Not.\ Roy.\ Astron.\ Soc.\  {\bf 416}, 3017 (2011)
  doi:10.1111/j.1365-2966.2011.19250.x
  [arXiv:1106.3366 [astro-ph.CO]].

\bibitem{Ross:2014qpa}
  A.~J.~Ross, L.~Samushia, C.~Howlett, W.~J.~Percival, A.~Burden and M.~Manera,
  Mon.\ Not.\ Roy.\ Astron.\ Soc.\  {\bf 449}, no. 1, 835 (2015)
  doi:10.1093/mnras/stv154
  [arXiv:1409.3242 [astro-ph.CO]].

\bibitem{Cuesta:2015mqa}
  A.~J.~Cuesta {\it et al.},
  Mon.\ Not.\ Roy.\ Astron.\ Soc.\  {\bf 457}, no. 2, 1770 (2016)
  doi:10.1093/mnras/stw066
  [arXiv:1509.06371 [astro-ph.CO]].

\bibitem{Gil-Marin:2015nqa}
  H.~Gil-Mar¨ªn {\it et al.},
  Mon.\ Not.\ Roy.\ Astron.\ Soc.\  {\bf 460}, no. 4, 4210 (2016)
  doi:10.1093/mnras/stw1264
  [arXiv:1509.06373 [astro-ph.CO]].





\end{thebibliography}
\end{document}